\documentclass[twocolumn,showpacs,preprintnumbers,amsmath,amssymb]{revtex4}
\usepackage{graphicx}% Include figure files
\usepackage{dcolumn}% Align table columns on decimal point
\usepackage{bm}% bold math
\include{def}
\def\Journal#1#2#3#4{{#1} {#2} (#4) #3}

\def\NPA{Nucl. Phys. A}

\def\PLB{Phys. Lett.  B}
\def\PRL{Phys. Rev. Lett.}
\def\PRC{Phys. Rev. C}

\def\MPL{Mod. Phys. Lett. A}
\def\EPJ{Eur. Phys. J. A}
\def\MPL{Mod.Phys.Lett.}
\def\FB{Fizika B}
\def\PA{Physica A}
\def\JSP{J. Stat. Physics}
\def\PLA{Phys. Lett.  A}
\begin{document}
\preprint{\it{Presented at the European Nuclear Physics Conference,
16-20 March 2009, Bochum, Germany}}
\title{Collective Phenomena in Heavy Ion Collisions}
\author{M. Petrovici}%
\email{mpetro@nipne.ro} 
 \author{A. Pop}
 \affiliation{National Institute of Physics and Nuclear Engineering\\
  P.O.Box MG-6, Bucharest-Magurele, Romania
}%%
\date{\today}
\begin{abstract}
A review of the main results of  detailed flow analysis in highly central and semi-central 
heavy ion collisions at SIS energies is presented in the first part of this paper.
The influence of the mass of the colliding nuclei and centrality on the collective expansion and
the information on the equation of state of compressed and hot baryonic matter is discussed. The second
part is dedicated to a similar type of analysis, based on the behaviour of the average transverse 
momentum as a function of mass of different hadrons, at the other extreme of energy range, where free 
baryonic fireballs are produced. Information on the partonic and hadronic expansion, temperature and 
degree of thermal equilibrium in p+p and Au+Au central collisions at 200 A$\cdot$GeV is presented.
\end{abstract}
\maketitle
\section{\label{sec:1}Introduction}
As far as the new states of matter are obtained in laboratory using heavy ion collisions, 
 it is mandatory to have as much as possible under control the finite size and dynamical 
 aspects. One of these is the dynamical evolution of the transient piece of matter
 produced in heavy ion collisions. 
 The present experimental evidence supports the existence of collective expansion caused by the 
 pressure gradients built-up during the first phase of the collision.
 Understanding this expansion one could aim to pin down the properties of the initial phase of
 matter, before expansion and at the same time, to extract information on in-medium
 effects, equation of state or phase transitions.
     
     Results of a detailed analysis of the expansion properties of hot and compressed 
baryonic matter at SIS energies in highly central and mid-central heavy ion collisions 
will be presented in Chapters I and II, respectively. Chapter III presents results of
a similar analysis in terms of collective expansion and temperature at RHIC energies, in 
the free baryonic sector. Preliminary results seem to support the possibility to disentangle 
between the contributions coming from partonic and hadronic phases and to access information on 
the degree of thermal equilibration of these two stages of expanding matter formed
in central collisions at 200 A$\cdot$GeV. Conclusions will be presented in Chapter IV.
\section{\label{sec:2}Highly central symmetric collisions}
For the highly central Au + Au collision at 150 A$\cdot$MeV incident energy \cite{jeong}, the mean 
kinetic energy of fragments divided by their mass, was observed to present a dependence as a
function of atomic charge which deviates from the one typical for a thermal plus Coulomb scenario.
This experimental trend agrees rather well with a thermal motion superimposed onto a collective 
velocity field which in 
the non-relativistic limit can be written as: 
\begin{equation}
<E_{kin}^{cm}> = \frac{1}{2}A\cdot m_0\cdot <\beta_{flow}^2> + \frac{3}{2}"T"
\end{equation}
\noindent
$\frac{1}{2}m_0\cdot <\beta_{flow}^2>$ is the flow energy per nucleon and 
$"$T$"$ stands for the effect of non-explicit treatment of Coulomb effects in the above expression. 
This leads systematically to an overestimated value of the real temperature T \cite{pog}.
   While this trend was qualitatively reproduced by the QMD microscopic transport model \cite{hart}, the 
absolute value of mean kinetic energy per nucleon is systematically underestimated for all fragments.
The experimental values were reproduced rather well by a hybrid model based on hydrodynamical 
expansion coupled with a statistical fragmentation at the break-up moment \cite{pet1}. This supports the model 
prediction that different species originate with different probabilities as a function of position 
within the fireball and time, heavier fragments being emitted later, at lower
temperature and lower expansion. This was confirmed in the meantime 
by small angle correlations studies for pairs
of nonidentical reaction species \cite{kotte},\cite{pet2}. A detailed analysis of experimental 
FOPI-Phase II data based on the equation above was performed for three symmetric systems 
(Au+Au, Xe + CsI, Ni + Ni), different incident energies and two regions of polar angles in the center 
of mass system for highly central collisions \cite{pet3},\cite{pet4}.      
\begin{figure}
\includegraphics[scale=0.37]{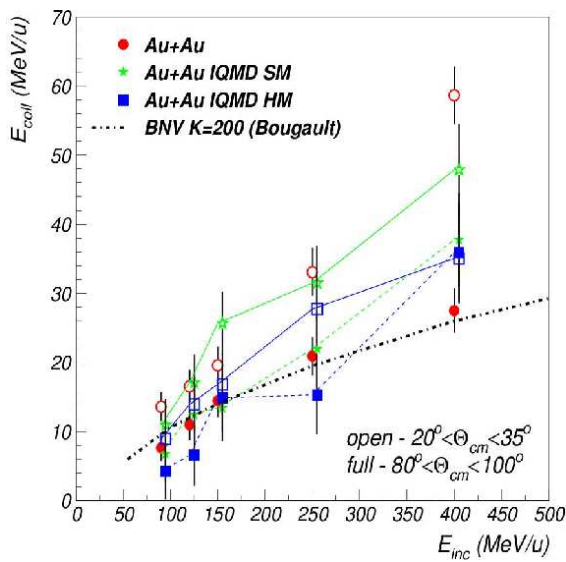}
\caption{\label{fig:Fig1}
Collective energy,
$E_{coll}$, as a function of incident energies in highly central Au+Au 
collisions (1\% total cross section). Open circles $25^{\circ}\le\theta_{cm}\le 45^{\circ}$,
full circles $80^{\circ}\le\theta_{cm}\le 100^{\circ}$. IQMD predictions are represented by 
open and full squares for hard equation of state (HM) and open and full stars for 
soft equation of state (SM), for the corresponding polar regions. BNV prediction using a soft 
equation of state is represented by dotted dashed line}
\end{figure}  
 While in the forward polar region, 
$25^{\circ}\le\theta_{cm}\le 45^{\circ}$, all three systems show the same flow at all measured energies,
along the transverse direction and mid-rapidity, i.e. $80^{\circ}\le\theta_{cm}\le 100^{\circ}$, the 
collective expansion is lower, its dependence on the incident energy is not that steep and at the same
incident energy the collective expansion increases with the mass of the colliding system.
Calculations based
on IQMD model at 250 A$\cdot$MeV showed that the nucleons emitted at $90^{\circ}$ suffer in the average 3.5
collisions relative to 1.7 of those emitted at smaller polar angles. This tells that at $90^{\circ}$
the observed effect is most probably coming from an equilibrated fireball, while at forward angles the
Corona effects and uncertainty in impact parameter selection play an important role. Experimental 
slopes obtained for highly central Au+Au collisions, 1\% total cross section, selected using the ratio
of transverse and longitudinal energies of all detected and identified charged particles
($E_{rat}=\sum_{i}E_{\perp,i}/\sum_{i}E_{\parallel,i}$ in the c.m. system) presented in
Fig.~\ref{fig:Fig1} \cite{stoi1}, support the microscopic transport model 
estimates, IQMD \cite{hart} and BNV \cite{bug} based 
on soft equation of state.
\section{\label{sec:3}Mid-central collisions}
\begin{figure}
\includegraphics[scale=0.35]{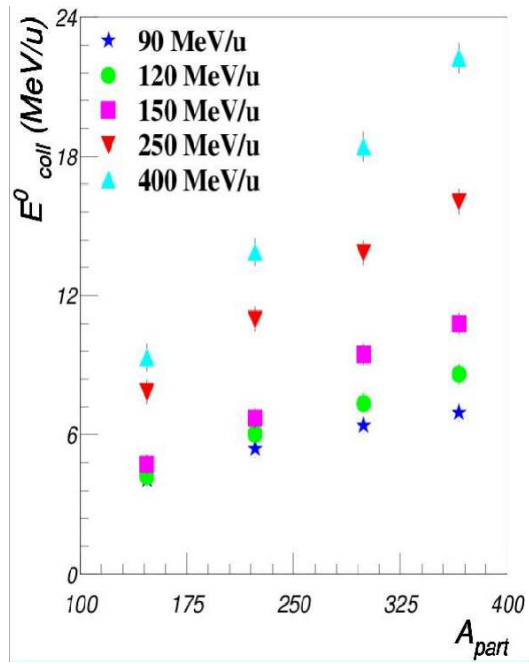}
\caption{\label{fig:Fig2}
Azimuthaly averaged $E_{coll}$ as a function of $A_{part}$ for Au+Au 
collisions at 90, 120, 150, 250 and 400 A$\cdot$MeV and
$80^{\circ}\le\theta_{cm}\le 100^{\circ}$.}
\end{figure}

\begin{figure} [ht]
\includegraphics[scale=0.35]{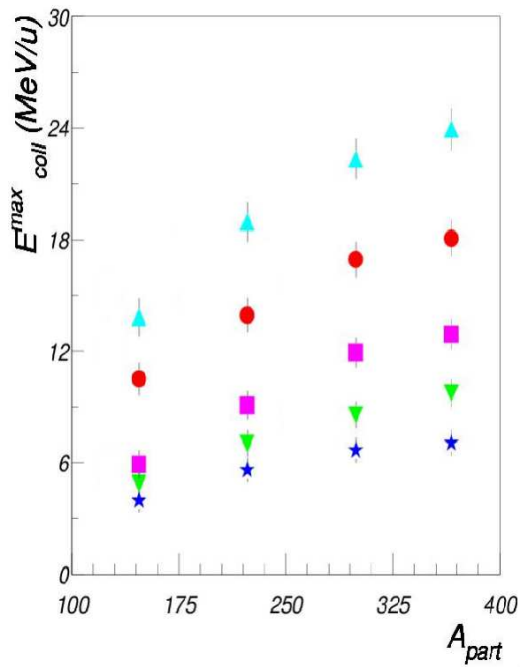}
\caption{\label{fig:Fig3}
$E_{coll}$ as a function of $A_{part}$ for Au+Au 
collisions at 90, 120, 150, 250 and 400 A$\cdot$MeV,
$80^{\circ}\le\theta_{cm}\le 100^{\circ}$ and
$72^{\circ}\le\phi\le 108^{\circ}$,$252^{\circ}\le\phi\le 288^{\circ}$ (same symbols as
in Fig.~\ref{fig:Fig2})}
\end{figure}

What else could one learn at these incident energies going to mid-central collisions ? As far as 
concerns the centrality dependence of collective expansion one has to be aware that an average on
azimuth (Fig.~\ref{fig:Fig2}) automatically includes contribution coming from shadowing effects due to spectator 
matter
while the values extracted at $90^{\circ}$ (Fig.~\ref{fig:Fig3}) are strongly influenced by the aspect ratio of the fireball.
 With this in mind, one can see  a systematic increase of flow energy with the centrality
($A_{part}$ estimated in a sharp cut-off geometrical model)
the effect being more pronounced at the higher measured incident energies.
 A detailed analysis of the collective expansion azimuthal distributions at mid-rapidity, different impact 
parameters and incident energies for Au + Au and Xe + CsI has been done by the FOPI Collaboration \cite{stoi2}.
 The qualitative agreement between the predictions of the model based on hydrodynamical expansion coupled 
with a statistical fragment formation at the break-up moment \cite{pet1}, under the hypothesis that the expansion starts at 
the maximum overlap and perfect shadowing of the spectator matter, let us to conclude that different regions 
of azimuth capture different periods of central fireball expansion \cite{pet2},\cite{stoi2}. If this is the 
case, then the amplitude of the azimuthal oscillation of $E_{coll}$ is recommended as a sensitive observable 
to the equation of state of hot and compressed baryonic matter produced at these energies.     
\begin{figure}
\includegraphics[scale=0.22]{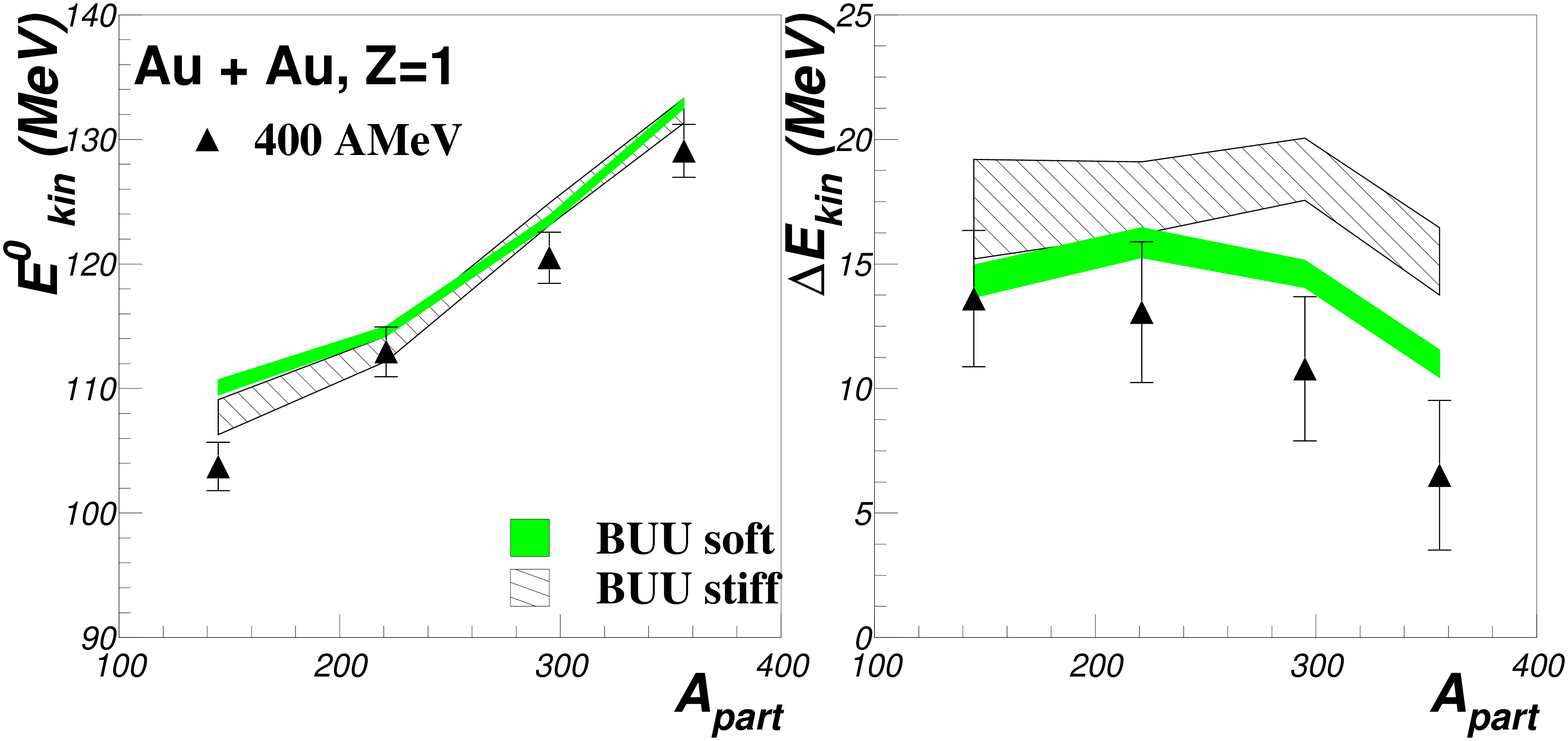}
\caption{\label{fig:Fig4}
$E_{kin}^{0}$ and ${\Delta}E_{kin}$
as a function of $A_{part}$, for Z=1 (A=1,2,3) fragments,
Au+Au at 400 A$\cdot$MeV. The experimental
results are represented by triangles, while the BUU results are
represented by gray zones for soft EoS and by dashed zones for stiff
EoS, respectively.}
\end{figure} 
 Transport model calculations based on BUU code \cite{dan}
using momentum dependent mean fields ($m^*$/m=0.79), in-medium
elastic cross sections
($\sigma = \sigma_0 \,
\mbox{tanh}(\sigma^{free}/\sigma_0)$ with $\sigma_0 =
\rho^{-2/3}$)
and soft (K=210 MeV, gray zone) or
stiff (K=380 MeV, dashed zone) EoS are compared in Fig.~\ref{fig:Fig4} 
and Fig.~\ref{fig:Fig5} with 
the experimental results. One should mention that the light fragments 
(up to A=3) are produced in a few-nucleon processes inverse to composite 
break-up, relative to the general coalescence recipe used by microscopic
transport codes.

The measured relative yields are nicely reproduced by this model, especially 
at higher incident energies \cite{pog}.
For comparison with the experiment, the 
calculated azimuthal distribution has been smeared according to the measured
reaction-plane dispersion
values. The azimuthal distribution of the average kinetic energy of
different species and of the collective energy per nucleon extracted from these were
fitted with: 
\begin{equation}
<E_{kin}>=E_{kin}^0-{\Delta}E_{kin}{\cdot}cos2{\Phi}
\end{equation} 
and
\begin{equation}
E_{coll}=E^0_{coll}-{\Delta}E_{coll}{\cdot}cos2{\Phi}
\end{equation}
respectively.
\begin{figure} [ht]
\includegraphics[scale=0.22]{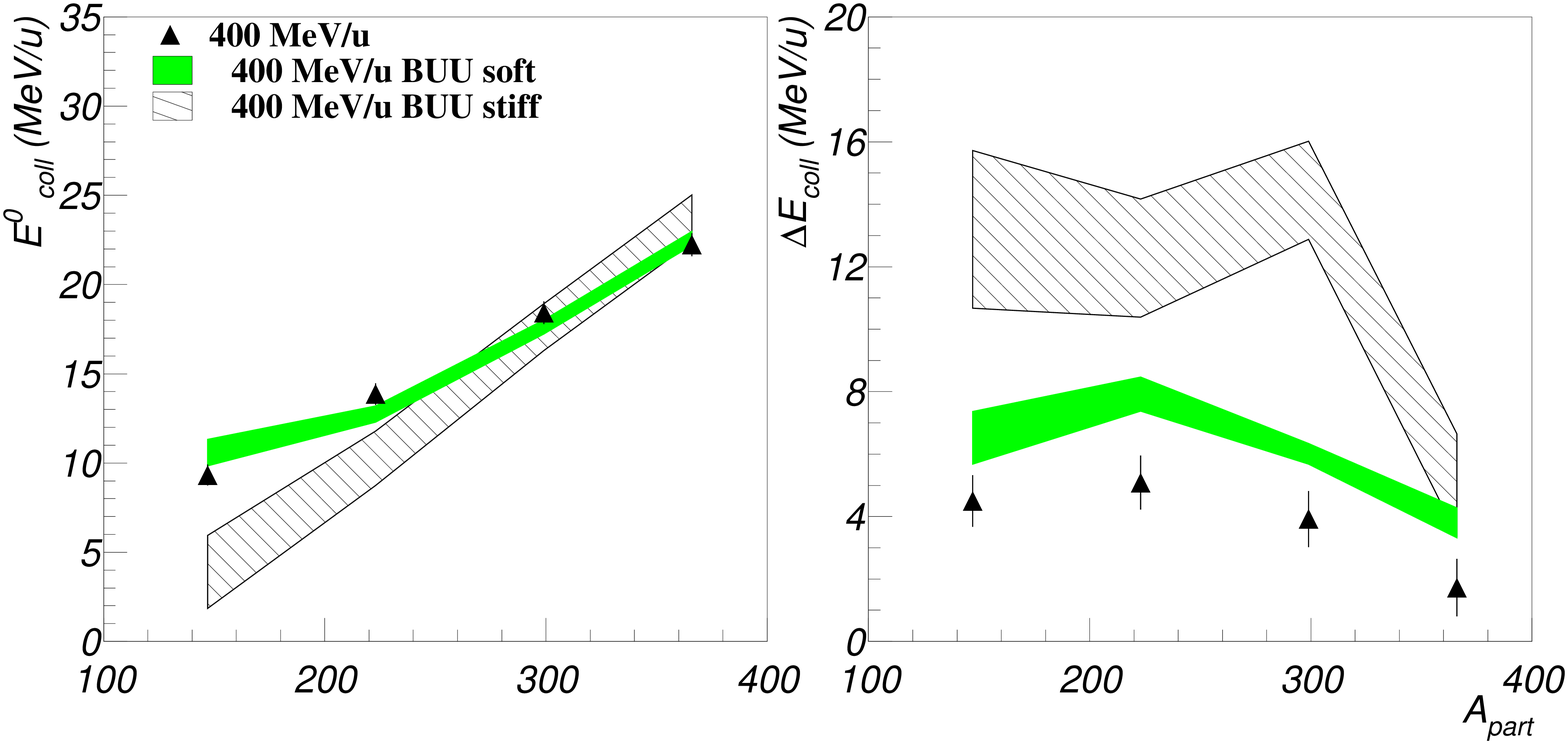}
\caption{\label{fig:Fig5}
$E_{coll}^{0}$ and ${\Delta}E_{coll}$
as a function of $A_{part}$, for
Au+Au at 400 A$\cdot$MeV. The symbols have the same interpretation as in
Fig.~\ref{fig:Fig4}.}
\end{figure}
As one could see in these figures, $E_{kin}^0$ for 
Z=1 (A=1,2,3) fragments (Fig.~\ref{fig:Fig4}) and $E_{coll}^0$ (Fig.~\ref{fig:Fig5}) are very little sensitive 
to the equation of state, both EoS parameterizations showing quite good 
agreement with the data. As far as concerns $\Delta E_{kin}$ and $\Delta E_{coll}$, 
the calculations with the soft EoS reproduce the overall trends of the experiment
while the calculations with the stiff EoS overestimate significantly the 
${\Delta}E_{kin}$ and ${\Delta}E_{coll}$ values 
at  higher and lower centralities, respectively. 
The results 
presented above support the conclusion that the equation of state of baryonic matter
at densities of about 2$\rho_0$ and temperatures of about 50-70 MeV is soft. This 
conclusion seems to be also supported by kaon production in heavy ion collisions
as it was shown by KAOS Collaboration \cite{sturm}.  
\section{\label{sec:4}Towards baryonic free matter}
\begin{figure} [ht]
\includegraphics[scale=0.145]{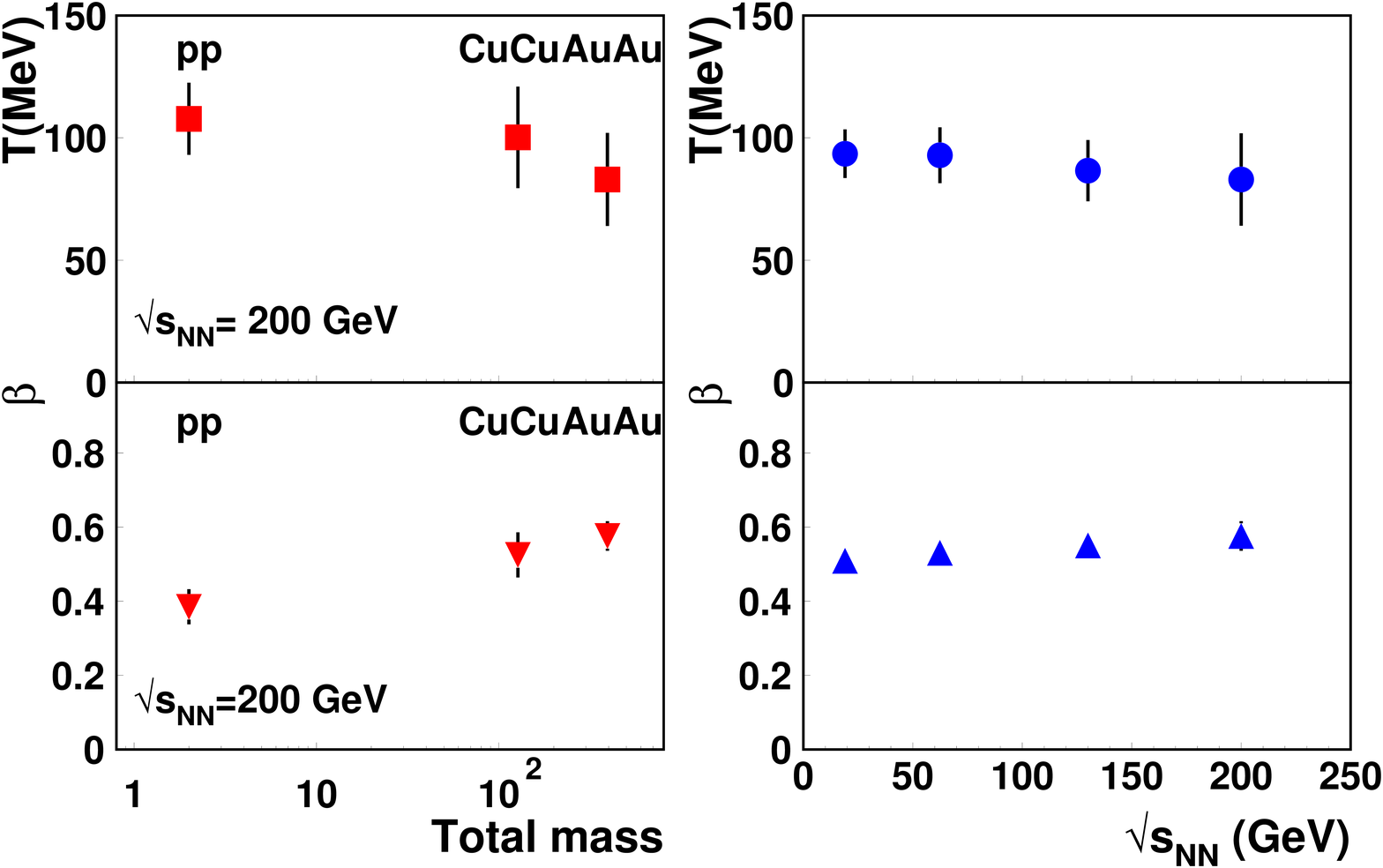}
\caption{\label{fig:Fig6}Left part: the temperature T and expansion velocity
$\beta$ as a function of total mass of the colliding system at 200 A$\cdot$GeV; Right part: 
the temperature and $\beta$ for Au + Au as a function of center of mass energy per nucleon.}
\end{figure}

The experimental information obtained at AGS, SPS and RHIC energies confirmed that 
the particle distributions reflect mainly the conditions reached by the system in 
its final state. Therefore, information originating from earlier stages is mandatory in 
order to conclude whether the system passed through a partonic phase and extract 
information on such a phase.
 In central collisions, the final collective transverse, azimuthally isotropic flow,
cumulates any collective flow generated during the evolution of the fireball, i.e. from 
the partonic and hadronic phases, the yield ratios reflecting mainly the statistical nature 
of the hadronization process.
 This was nicely shown at RHIC, by the STAR Collaboration \cite{olga}, by analyzing the 
transverse momentum distribution for common and rare particles using the blast wave model.
The extracted kinetic freeze-out temperature and the flow velocity show a clear dependence
as a function of centrality while the chemical freeze-out temperature, extracted from particle 
yield ratios using statistical model stays constant, at about 170 MeV. At the same time
the data seem to indicate a sequential freeze-out of particles, hyperons decoupling from the 
system earlier at temperatures closer to the chemical temperature value.
 As we have seen in the first part of the present paper, a useful observable for a detailed study 
of collective flow would be the mean kinetic energy as a function of mass of different species.
 In the following we shall present the results of a similar analysis, this time in terms of 
the average transverse momentum $\langle p_t \rangle$ as a function of hadron mass for different colliding 
systems and energies.
 A careful examination of the experimental $\langle p_t \rangle$ values  \cite{abel} as a function of mass for
${\pi}^{\pm}$, $K^{\pm}$, p and $\bar{p}$ evidences that the slope 
increases going from p+p to Cu+Cu and
Au+Au at the same incident energy, i.e. 200 A$\cdot$GeV. Although less pronounced, similar trend is observed 
as a function of incident energy for a given system \cite{petpop}. In order to extract quantitative 
information we used the blast wave model for 
calculating the $\langle p_t \rangle$ as a function of mass:
\begin{equation}
<p_t> =
\frac{\int_{0}^{\infty} p_t^2 f(p_t)dp_t}
{\int_{0}^{\infty} p_t f(p_t)dp_t}
\end{equation}  
where:
\begin{equation}
f(p_t)\sim\int_0^R rdrm_t I_0\left(\frac{p_tsinh\rho}{T}\right)K_1\left(\frac{m_tcosh\rho}{T}\right)
\end{equation}
\begin{equation}
\rho  = tanh^{-1}{\beta}_r
\end{equation} 
and
\begin{equation}
{\beta}_r(r) = \beta_s\left(\frac{r}{R}\right)
\end{equation} 
 The temperature T and expansion velocity $\beta_s$ were the free parameters used to fit the 
experimental $\langle p_t \rangle$ as a function of mass \cite{schneder}. Within this ansatz
$\beta=\frac{2}{3}\beta_{s}$.
 The results are presented in Fig.~\ref{fig:Fig6}. 
 The left part shows the temperature and expansion velocity
$\beta$ as a function of total mass of the colliding system at 200 A$\cdot$GeV and the right part 
the temperature and $\beta$ for Au + Au as a function of center of mass energy per nucleon.
The temperature drops by $\sim$ 20 MeV and the expansion velocity is increasing from p+p towards heavy systems.
 Although the 
decrease in temperature is moderate it is consistent with the interpretation that stronger expansion 
observed in heavier combinations cools down the fireball and therefore the final, kinetic break-up
temperature is lower.
\begin{figure}
\includegraphics[scale=0.4]{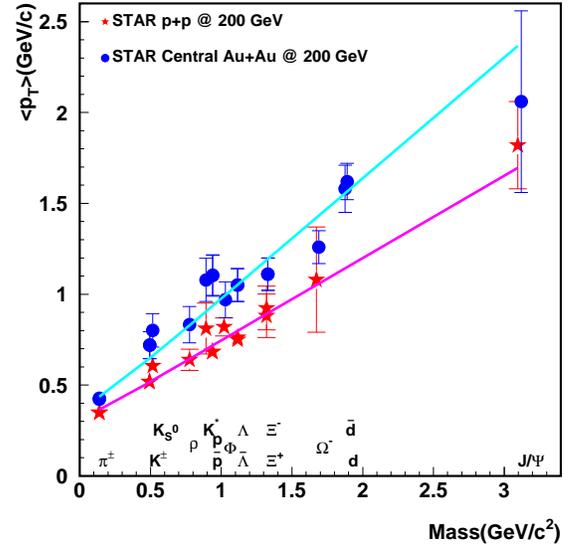}
\caption{\label{fig:Fig7}Mid rapidity average transverse momentum as a function of mass of different 
particles
measured by STAR Collaboration.}
\end{figure}
 Let's look now to $\langle p_t \rangle$ as a function of mass for all particles measured by the STAR 
Collaboration at 200 A$\cdot$GeV for p+p and Au+Au collisions \cite{abelev} 
presented in Fig.~\ref{fig:Fig7}.  
 The two lines represent the results of the fits using the above expressions for 
p+p ( dark line)  and Au+Au (gray line). The corresponding temperatures and 
$\beta$ values are: T=111.6$\pm$23.8 MeV and $\beta$=0.39$\pm$0.06 for p+p and 
T=109.8$\pm$16.5 MeV and $\beta$=0.50$\pm$0.04 for Au+Au. 
While the freeze-out kinetic 
temperature seems to be similar for p+p and Au+Au, the expansion is much more
violent in the Au+Au case. If for Au+Au one considers only $\pi^{\pm}$, $K^{\pm}$, $K^{\ast}$,
$K_s^0$, p, $\bar{p}$, d, $\bar{d}$ particles, the obtained temperature is T=98.7$\pm$19.5 MeV 
and $\beta$=0.54$\pm$0.04. 
 It is well known that for understanding the particle production in high energy and 
nuclear physics, many authors used Tsallis statistics \cite{tsallis}. Tsallis' 
generalization of Boltzmann-Gibbs extensive statistics, based on the definition
of a q-deformed entropy functional, is supposed to be adequate for describing 
systems characterized by memory effects and long range-interactions. 
The Boltzmann-Gibbs
statistics is recovered in the limit q$\rightarrow$1, therefore (q-1) is interpreted as 
a measure of the degree of non-equilibrium, the temperature T being interpreted as
the average temperature. If we replace in Eq. (4), f($p_t$) with the distribution 
corresponding to the blast wave model in which the Tsallis  statistics has been implemented \cite{lav}, \cite{tang}: 
\begin{widetext}
\begin{equation}
f(p_t) = m_T 
\int_{-Y}^Y cosh(y)dy\int_{-\pi}^{\pi}d{\phi}\int_0^Rrdr(1+\frac{q-1}{T}
(m_Tcosh(y)cosh(\rho)-p_tsinh(\rho)cos(\phi)))^{-1/(q-1)}
\end{equation} 
\end{widetext}
and fit the same experimental data with the new expression, the results 
presented in Fig.~\ref{fig:Fig8} are obtained.
\begin{figure} [ht] 
\includegraphics[scale=0.095]{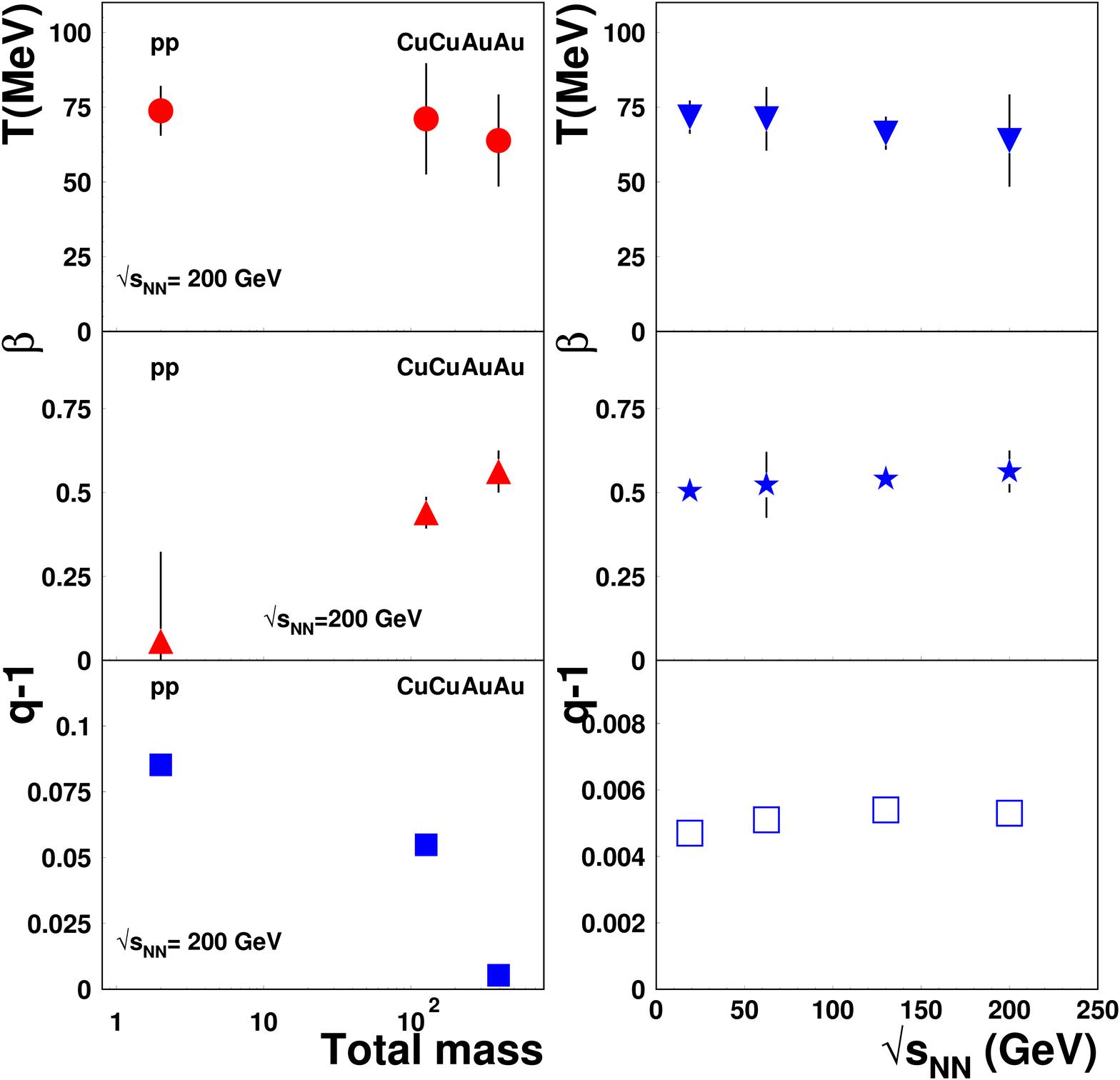}
\caption{\label{fig:Fig8}Left part: the temperature T, expansion velocity
$\beta$ and q parameters as a function of total mass of the colliding system at 200 A$\cdot$GeV; 
Right part: 
the temperature, $\beta$ and (q-1) parameters for Au + Au as a function of center of mass energy per 
nucleon.}
\end{figure}  
\begin{table} [h]
\caption{\label{tab:Table1}}
\begin{tabular}{|c|c|c|c|c|} \hline
System & p + p  & p + p & Au + Au & Au + Au \\ \hline
Model & BGBW  & TBW & BGBW & TBW \\ \hline
T& 111.6$\pm$23.8 & 78.86$\pm$10.13 & 109.8$\pm$16.5 & 86.8 $\pm$1.54 \\ \hline
$\beta$& 0.39$\pm$0.06 & 0.027$\pm$0.10 & 0.50$\pm$0.04 & 0.48$\pm$0.04 \\ \hline
q& 1.0 & 1.087$\pm$0.002 & 1.0 & 1.025$\pm$0.003 \\ \hline
\end{tabular}
\end{table}
 For p+p the boosted Tsallis scenario gives a negligible 
expansion velocity.
 As far as concerns the q parameter, one could observe a strong decrease as a function of
total mass of the colliding system, large deviation from Boltzmann statistics, 
(q-1)$\simeq$0.085, for p+p to relative small value, 0.005, for Au+Au being observed. The 
value for Au+Au stays almost constant, $\simeq$0.005 as a function of incident energy.
 One should mention that a similar trend as a function of incident energy but for larger 
(q-1) values was observed in $e^+e^-$ collision, obtained from analyzing the $p_t$ spectra in 
terms of simple, not boosted, Tsallis distribution \cite{bediaga}.

 If all species are considered, as in Fig.~\ref{fig:Fig7}, the results of the fit are 
 summarized in Table~\ref{tab:Table1} where the 
  blast wave models with Boltzmann-Gibbs (BGBW) and respectively Tsallis (TBW) statistics are compared. 
$\beta$ is almost
zero, therefore no expansion is built up in p+p collision at 200 A$\cdot$GeV and the degree of 
non-equilibrium is about a factor of four larger in p+p relative to Au+Au.
For Au+Au the extracted 
temperature is about 20 MeV lower and $\beta$ is within the error bars similar with the one obtained 
using the blast wave model with Boltzmann statistics scenario. If one considers separately common particles, i.e. hadrons with 
larger interaction cross section, the results are presented in Table~\ref{tab:Table2}. 
\begin{table} [h]
\caption{\label{tab:Table2}}
\begin{tabular}{|c|c|c|} \hline
& Au + Au & Au + Au \\ \hline
&BGBW  & TBW \\ \hline
T [MeV]& 98.7$\pm$19.5 & 79.05$\pm$0.04 \\ \hline
$\beta$& 0.54$\pm$0.04 & 0.53$\pm$0.0005 \\ \hline
q& 1.0 & 1.0175$\pm$0.0018\\ \hline
\end{tabular}
\end{table}
The temperature decreases, 
$\beta$ slightly increases and the deviation from Boltzmann statistics is reduced. If we consider only 
the hyperons and J/$\Psi$, as it is shown in Table~\ref{tab:Table3}, a temperature around 200 MeV, 
$\beta \simeq$0.3 and larger deviation from global equilibrium is obtained.    
\begin{table} [h]
\caption{\label{tab:Table3}}
\begin{tabular}{|c|c|} \hline
& Au + Au \\ \hline
& TBW \\ \hline
T [MeV]& 198.0$\pm$7.6 \\ \hline
$\beta$& 0.32$\pm$0.012 \\ \hline
q& 1.0247$\pm$0.0043\\ \hline
\end{tabular}
\end{table}
This suggests that strange  and heavy flavour hadrons keep the characteristics of expansion 
at the hadronization moment, characterized by lower values of $\beta$, higher temperature
and based on Tsallis statistics interpretation, not fully equilibrated. 
\begin{figure}  [ht]
\includegraphics[scale=0.106]{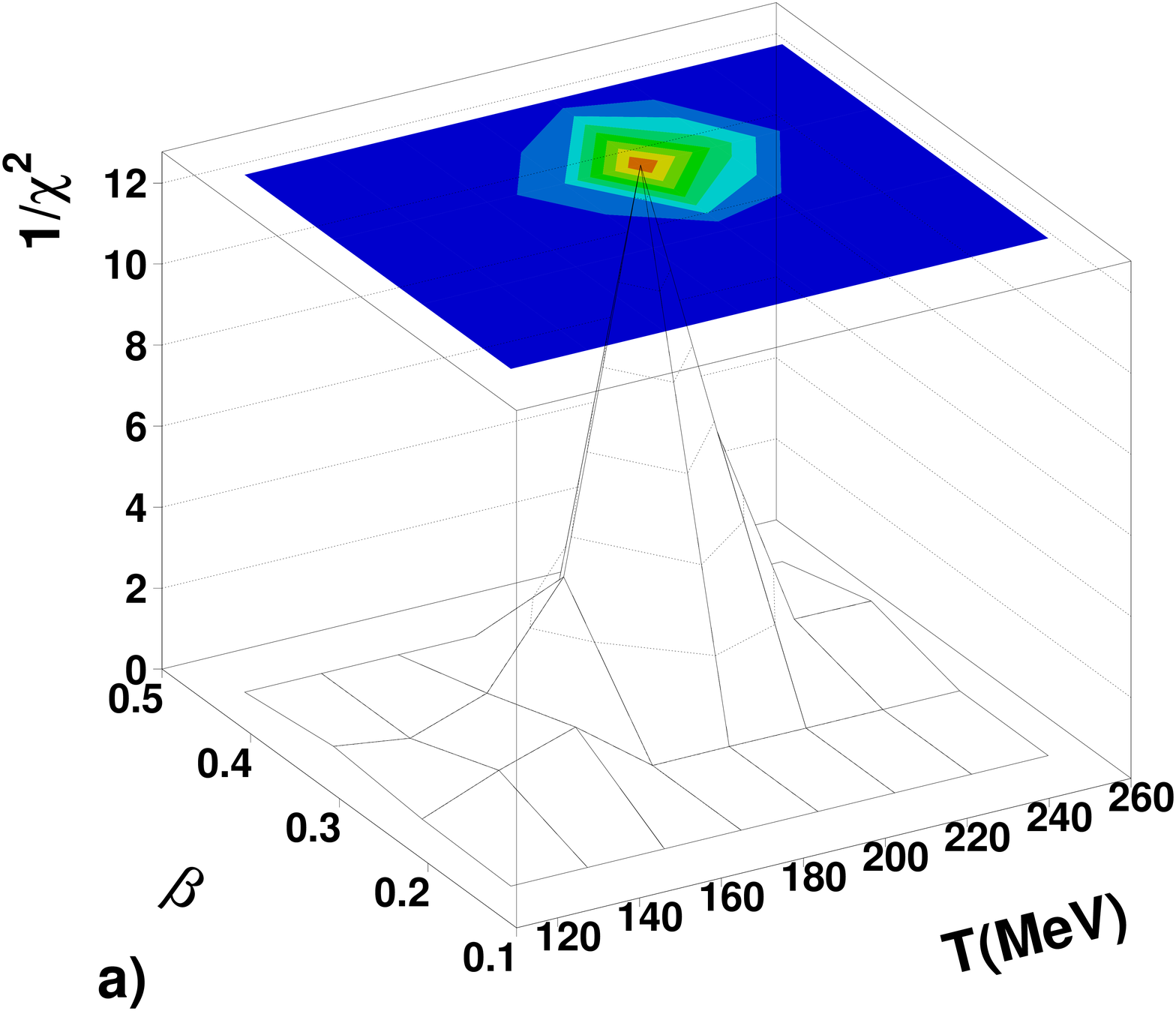}
\includegraphics[scale=0.106]{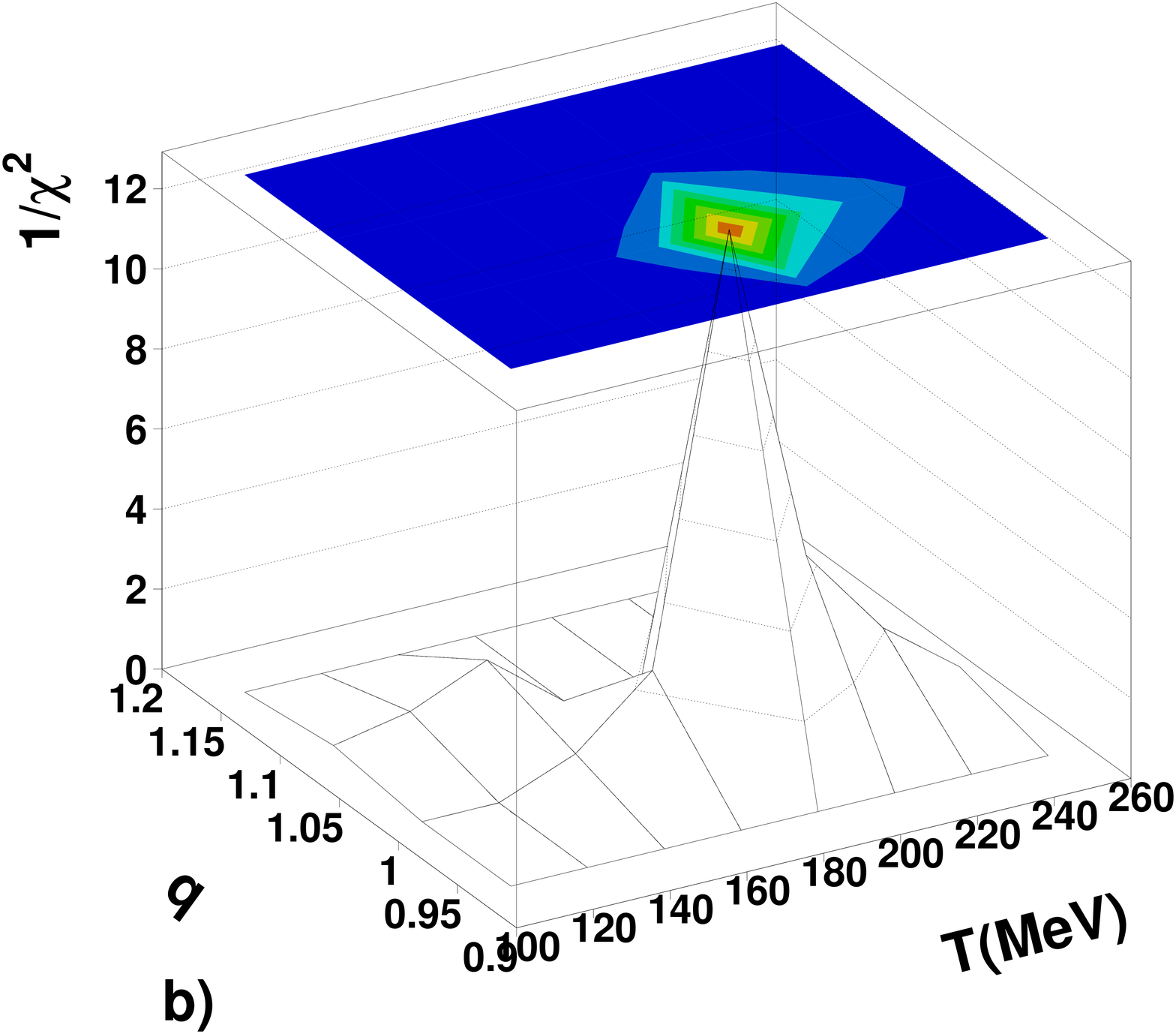}
\caption{\label{fig:Fig910}a) 1/$\chi^2$ as a function of temperature and $\beta$;
b)
1/$\chi^2$ as a function of temperature and q for the result of the fit 
listed in Table III.}
\end{figure}    
 After hadronization, strongly interacting particles, within still highly dense environment 
continue to build up expansion, cooling down the system and approaching a global equilibrium.

The Minuit package was used to perform the least $\chi^2$ fit. The quality of the fit
was reasonably good as can be seen in Fig.~\ref{fig:Fig910}, where 1/$\chi^2$ is represented as a function
of two pairs of parameters for the result of the fit listed in Table III.
\section{\label{sec:5}Conclusions}
In this paper we tried to present a short review of the main results of detailed flow 
analysis at incident energies where compressed and hot baryonic fireballs are 
produced. The experimental data support model predictions that different species originate 
with different probabilities as a function of position 
within the fireball and time, heavier fragments being emitted later, at lower
temperature and lower expansion. The best agreement between experiment and theoretical estimates
based on microscopic transport models is obtained if a soft equation of state is considered.

 Similar type of analysis extended at the other extreme of incident energies, where free 
baryonic matter is produced, shows that for central collisions, azimuthally isotropic flow 
of different species could be used to pin down the relative contributions coming from partonic
and hadronic levels. 
The kinetic temperature and expansion velocity extracted from experimental
$<p_t>$ dependence on the hadron mass using boosted Boltzmann and Tsallis expressions 
support the conclusion that strange and heavy flavour hadrons carry the information
from the partonic stage characterized by higher temperature and lower expansion. The results
based on boosted Tsallis distribution (within the limits of its applicability and interpretation) 
indicate high degree of non-equilibrium and
 missing expansion in pp collisions at RHIC energy. For the Au+Au collision similar analysis for
 strange and
 heavy flavour hadrons gives a higher temperature, less violent expansion and 
 higher degree of 
 non-equilibrium of the initial, deconfined  
matter produced at the highest RHIC energy. Common hadrons, strongly interacting after
hadronization, continue to build up expansion, cool down the system and bring it towards 
 a global equilibrium.
 
 It is a real challenge for the near future experiments at LHC to use the potentiality of flow
phenomena to extract new physics in an energy domain of about 30 times larger than the one attainable
at RHIC. Preliminary results based on Monte Carlo simulations show that the ALICE experiment is 
able to reconstruct with high accuracy the $p_t$ distribution for most of the particles making
the type of analysis presented in this paper feasible and promising.  
 
\end{document}